\documentclass[12pt]{article}
\usepackage{epsfig}

\parskip=\medskipamount % space between paragraphs (TeX)
plus.3em minus.5em \abovedisplayshortskip=.5em plus.2em minus.4em
\renewcommand{\thesection}{\arabic{section}}

\catcode`@=11

\@addtoreset{equation}{section} \@addtoreset{equation}{subsection}
\def\theequation{\ifnum\value{section}=0 \arabic{equation}\ignorespaces
\else \ifnum\value{section}=-1 A.\arabic{equation}\ignorespaces
\else \ifnum\value{subsection}=0
\thesection.\arabic{equation}\ignorespaces \else
\thesection.\arabic{subsection}.\arabic{equation}\ignorespaces
                             \fi
                        \fi
                   \fi}

{\catcode`\'=\active \def'{{}^\bgroup\prim@s}}

\catcode`@=12

\newcommand{\bq}{\begin{equation}}
\newcommand{\be}{\begin{equation}}
\newcommand{\fq}{\end{equation}}
\newcommand{\ee}{\end{equation}}
\newcommand{\bqr}{\begin{eqnarray}}
\newcommand{\beqs}{\begin{eqnarray}}
\newcommand{\fqr}{\end{eqnarray}}
\newcommand{\eeqs}{\end{eqnarray}}

\def\bop#1{\setbox0=\hbox{$#1M$}\mkern1.5mu
    \vbox{\hrule height0pt depth.04\ht0
    \hbox{\vrule width.04\ht0 height.9\ht0 \kern.9\ht0
    \vrule width.04\ht0}\hrule height.04\ht0}\mkern1.5mu}
\begin{document}
\thispagestyle{empty}

\vskip .6in
\begin{center}

{\bf Computational Improvements to Matrix Operations}

\vskip .6in

{\bf Gordon Chalmers}
\\[5mm]
% {\em address \\
%      address \\
% Los Angeles, CA } \\

{e-mail: gordon@quartz.shango.com}

\vskip .5in minus .2in

{\bf Abstract}

\end{center}

An alternative to the matrix inverse procedure is presented.  Given a bit 
register which is arbitrarily large, the matrix inverse to an arbitrarily 
large matrix can be peformed in ${\cal O}(N^2)$ operations, and to matrix 
multiplication on a vector in ${\cal O}(N)$.  This is in contrast to the 
usual ${\cal O}(N^3)$ and ${\cal O}(N^2)$.  A finite size bit register can 
lead to speeds up of an order of magnitude in large matrices such as 
$500\times 500$.   The FFT can be improved from ${\cal O}(N\ln N)$ to 
${\cal O}(N)$ steps, or even fewer steps in a modified butterfly 
configuration.  

\vfill\break

The matrix inverse is the backbone to the modern STAP process; this has to 
be performed every time a set of data is trained in the field of view.  The 
complexity is one of the primary origins of the cost to the large computing 
required to process the data.  Any improvement in the complexity to performing 
the matrix inverse is a desirable.   

Typically the matrix inverse to $M_{N\times N}$ is performed in ${\cal O}(N^3)$ 
operations.  There are several variants, including the LU and QR variants.  The 
LU diagonalization is twice as fast as the (Guass) QR form \cite{GL}.   
The LU factorization requires splitting the matrix into the product of upper and 
lower diagonal matrices $M=LU$.  The inverse is performed by inverting the respective 
components.  The QR form requires splitting the matrix $M$ into an orthogonal 
component $Q$ times its projection $R$.  

There is a simplification of the matrix inverse by grouping the entries 
of the matrix $M_{ij}$ into larger numbers.  For example, the first row of the 
matrix has elements $N_1$, $N_2$, $\ldots$.  A larger number can be built of 
these entries by placing the digits into one number $N_1 N_2 \ldots$.  For the 
computing purposes a zero number $N_0$ with as many digits as the entries $M_{ij}$ 
is required, and the rows are grouped into the number $N_1 N_0 N_2 N_0 \ldots$.  
The rows of the matrix are now used as a single number in the diagonalization 
procedure in the LU factorization.  

For example, the zeroing out of the matrice's first column requires using the 
first row; a number $b_{i1}$ is used to multiply $M_{1j}$ so that 
$M_{i1}=-b_{i1} M_{i1}/M_{11}$.  This number multiplies the entire row of the 
matrix $M_{1j}$ and is added to the $i$th row.  In doing so, a set of zeroes is 
produced in the first column of the matrix $M$; the numbers $b_{i1}$ are placed 
in the lower diagonal factor matrix of $L$.  The procedure is iterated using the 
diagonal elements $M_{jj}$ to construct upper diagonal and lower matrices $L$ and 
$U$.  The computational cost of one of the multiplications is $N^2$ due to the 
$N$ elements in the row and the multiplications and additions to the $N$ 
elements in the column.  

In using the larger number the $N^2$ operations to create a zero column can be 
reduced to $N$ operations.  This requires the zero number $N_0$ to have a sufficient 
number of digits so that 

\bqr 
b (N_1 N_0 N_2 N_0 \ldots) = (b N_1) N_0 (b N_2) N_0 \ldots 
\fqr 
\bqr 
(N_1 N_0 N_2 N_0 \ldots) + (M_1 M_0 M_2 M_0 \ldots) = 
 (N_1+M_1) N_0 (N_2+M_2) N_0 \ldots \ , 
\fqr 
as one number.  The bit register in the processor has to be able to handle these 
two operations, multiplication by a scalar and addition.  The $N$ operations in 
the bit register to treat the multiplication and addition of the original row has 
been reduced to one multiplication and one addition.  The numbers $b$ which multiply 
the larger numbers $N_1 N_0 N_2 N_0 \ldots$ are collected into the lower diagonal 
matrix $L$.  

A separate matrix is required to discern if the subtraction of a positive  
number to this number is negative or positive.  For example, 

\bqr 
(N_1 N_0 N_2 N_0 \ldots) - (M_1 M_0 M_2 M_0 \ldots) 
\fqr 
could have negative entries $N_i-M_i$ but the absolute value is used in the 
composition of the number $N-M$.  The subtraction process doesnt work well 
in the procedure, and the numbers are separated into $N_i-M_i$ independently.  

The processing of using the larger numbers instead of the smaller numbers is 
that typical processes such as the matrix inverse and the FFT can be reduced 
in complexity from ${\cal O}(N^3)$ and $N \ln N$ to ${\cal O}(N^2)$ and $N$. 

\vskip .2in 
\noindent {\it STAP Example}  

The use of spacetime adaptive processing requires the training of data using a 
covariance matrix.  This matrix is canonically symmetric in the acquisition of 
data, satisfying the multiplicative product $X=x_i x_j$.  The inverse of the 
covariance matrix is unwieldy, being performed in ${\cal O}(N^3)$ steps, but 
must be performed in conventional STAP processes and signal location.  

The product $x_i x_j$ represents a probability distribution, with positive 
entries.  An alternative matrix satisifies $X_{ij}=-X_{ij}$ with positive 
entries along the diagonal, is far more convenient in the matrix inverse 
procedure.  The inverse of the latter matrix can be performed theoretically 
in ${\cal O}(N^2)$ steps.  The limitation is set by the number of bits in 
the bit register; the ${\cal O}(N^2)$ arises from an arbitrarily large bit 
register.  

Consider the LU reduction of the alternative covariance matrix ${\tilde X}$.  
The process of adding the rows to null the lower left triangular portion of 
U requires only adding the numbers $M_1/N_1*(N_1 N_0 N_2 N_0 \ldots)$ to  
$(M_1 M_0 M_2 M_0 \ldots)$.  For example the second row modeled by the single 
number $(M_1 M_0 M_2 M_0 \ldots)$ has a negative entry for $M_1$ and positive 
entries $M_j$.  The addition of $M_1/N_1*N_j$ to the $M_j$ occurs in one 
operation theoretically due to the size of the individual numbers.  The 
$M_1/N_1$ is stored in the lower left triangular matrix L in the process of 
the LU factorization.  The operation is repeated to first nullify the left 
column of U (except the diagonal component), and then the process is repeated 
for the other columns.  As there are $N^2$ components in the matrix the 
LU factorization of ${\tilde X}$ requires $N^2$ steps; this is considerably 
faster than $N^3$ when $N$ is of the order of a thousand or more.  

The question is whether data can be trained with the alternate to the covariance 
matrix.  The ${\tilde X}$ contains the same information but with minus signs 
placed to partially antisymmetrize.  It appears clear by the conventional 
use of stap in locating signals, and in eliminating noise, that this should 
be possible.

\vskip .2in 
\noindent{\it Matrix Multiplication}  

The same use of the bit register and organizing the rows of the matrix 
in terms of whole numbers can be used to simplify matrix multiplication.  
Usual multiplication of a matrix by a vector requires ${\cal O}(N^2)$ steps.  
This can be reduced to ${\cal O}(N)$ with a reordering of the matrix 
and vector information.

Consider all positive entries in the matrix $M$ and all positive entries 
in the vector $v$.  The vector consists of one number $(v_1 v_0 v_2 v_0 \ldots)$, 
and the columns of the matrix consist of individual numbers ${\tilde M}_j=
(M_1 M_0 M_2 M_0 
\ldots)$.  The multiplication of one column by the vector element is accomplished 
in one step: $v_i*{\tilde M}_j$, with the element of the vector used.  This results in 
$v_i {\tilde M}_j=(v_i M_1 M_0 v_i M_2 M_0 \ldots$.  The total matrix multiplication 
is then accomplished by adding the previous multiplications: $\sum v_i {\tilde M_j}$.  
The vector resultant from the matrix multiplications are stored in the decomposition 
of $\sum v_i {\tilde M_j}$.  The total operations to matrix multiply is $2N$ steps, 
and not ${\cal O}(N^2)$.  This reduction can be substantial for numbers $N$ of the 
order of a thousand.  

The previous example pertains to the matrix $M$ and vector consisting of positive 
values.  Minus signs in the matrix can be incorporated very simply by separating 
the elements ${\tilde M}_j=(M_1 M_0 M_2 M_0 \ldots)$ into the respective positive 
entries and negative entries: ${\tilde M}_j^+$ and ${\tilde M}_j^-$.  A vector with 
positive entries can be used to multiply both the $v_i {\tilde M}_j=v_i {\tilde M}_j^+ + 
v_i {\tilde M}_j^-$ entries.  The addition of the column vectors is achieved by 
adding separately the positive entries and the negative entries: $\sum v_i 
{\tilde M}_j^+$ 
and $\sum v_i {\tilde M}_j^-$.  Then the individual entries of the two terms 
are required to be subtracted.  The net total number of steps is ${\cal O}(N)$.  

A simple application of the of the matrix multiplication of a vector is the fast 
fourier transform.  The butterfly reduction of the usual multiplication of the vector 
lowers the ${\cal O}(N^2)$ to ${\cal O}(N \ln N)$ steps.  Theoretically, by separating 
the matrix into real and complex parts, with the minus signs handled separately, can 
achieve a theoretical FFT in ${\cal O}(N)$ steps.  This exponentially faster than 
the butterfly configuration.   

The butterfly configuration can also be analyzed with subtle memory allocation 
of the data transfer to an approximate $\ln N$ operations.  The data has to be 
reorderd in traversing the butterfly.

\vskip .2in 
\noindent{\it Conclusions}

The theoretical improvements in the matrix inverse from ${\cal O}(N^3)$ steps 
to ${\cal O}(N^2)$ steps, and matrix times vector from ${\cal O}(N^2)$ to ${\cal 
O}(N)$ steps has profound impact in computational science.  Unfortunately, a 
bit register of a large size is required.  In conventional computing registers, 
there is a waste depending on the data size.  For example, a 256 bit register 
handling 32 bit data can be optimized by a factor of 8, which is still substantial.  

Ideally, the theoretical drop of the matrix inverse and the matrix multiplication 
by a factor of $N$ is suitable for more advanced computing apparatus.  The 
theoretical bounds in the optimization are achieved with large bit registers, 
which can be designed in several contexts.

\vfill\break

\end{document}